\documentclass[sigconf]{acmart}

\copyrightyear{2018} 
\acmYear{2018} 
\setcopyright{acmcopyright}
\acmConference[KDD '18]{The 24th ACM SIGKDD International Conference on Knowledge Discovery & Data Mining}{August 19--23, 2018}{London, United Kingdom}
\acmBooktitle{KDD '18: The 24th ACM SIGKDD International Conference on Knowledge Discovery & Data Mining, August 19--23, 2018, London, United Kingdom}
\acmPrice{15.00}
\acmDOI{10.1145/3219819.3219869}
\acmISBN{978-1-4503-5552-0/18/08}

\usepackage{amsmath}
\usepackage{graphicx}
\usepackage{color}
\usepackage[export]{adjustbox}
\usepackage[caption=false]{subfig}
\usepackage{tabularx}
\usepackage{booktabs}
\usepackage{multirow}
\usepackage{array}
\usepackage{enumitem}
\usepackage{url}
\usepackage{hyperref}

\usepackage{algorithm}
\usepackage{algorithmicx}
\usepackage{algpseudocode}
\algnewcommand\algorithmicinput{\textbf{INPUT:}}
\algnewcommand\INPUT{\item[\algorithmicinput]}
\algnewcommand\algorithmicoutput{\textbf{OUTPUT:}}
\algnewcommand\OUTPUT{\item[\algorithmicoutput]}

\usepackage{amsmath}
\DeclareMathOperator*{\minimize}{minimize}
 
\graphicspath{ {img/} }

\def\etal{~et~al. }

\def\bA{\textbf{A}}

\def\bH{\mathbf{H}}
\def\bM{\mathbf{M}}

\def\bW{\textbf{W}}
\def\bZ{\textbf{Z}}

\def\cG{\mathcal{G}}
\def\cV{\mathcal{V}}
\def\cE{\mathcal{E}}

\usepackage{array}
\newcolumntype{L}[1]{>{\raggedright\let\newline\\\arraybackslash\hspace{0pt}}m{#1}}
\newcolumntype{C}[1]{>{\centering\let\newline  \\\arraybackslash\hspace{0pt}}m{#1}}
\newcolumntype{R}[1]{>{\raggedleft\let\newline \\\arraybackslash\hspace{0pt}}m{#1}}

\begin{document}
\title{Billion-scale Commodity Embedding for E-commerce Recommendation in Alibaba}
\author{Jizhe Wang, Pipei Huang\footnotemark[1]}
\affiliation{
  \institution{Alibaba Group}
  \city{Hangzhou and Beijing}
  \country{China}
}\email{{jizhe.wjz, pipei.hpp}@alibaba-inc.com}
\email{{jizhe.wjz, pipei.hpp}@gmail.com}

\author{Huan Zhao}
\affiliation{%
	\institution{Department of Computer Science and Engineering}
	\institution{Hong Kong University of Science and Technology}
	\city{Kowloon}
	\country{Hong Kong}
}\email{hzhaoaf@cse.ust.hk}

\author{Zhibo Zhang, Binqiang Zhao}
\affiliation{
  \institution{Alibaba Group}
  \city{Beijing}
  \country{China}
}\email{{shaobo.zzb, binqiang.zhao}@alibaba-inc.com}

\author{Dik Lun Lee}
\affiliation{%
	\institution{Department of Computer Science and Engineering}
	\institution{Hong Kong University of Science and Technology}
	\city{Kowloon}
	\country{Hong Kong}
}
\email{dlee@cse.ust.hk}

\begin{abstract}

\renewcommand{\thefootnote}{\fnsymbol{footnote}}
\footnotetext[1]{Pipei Huang is the Corresponding author.}

Recommender systems (RSs) have been the most important technology for increasing the business in Taobao, the largest online consumer-to-consumer (C2C) platform in China. There are three major challenges facing RS in Taobao: scalability, sparsity and cold start. In this paper, we present our technical solutions to address these three challenges. The methods are based on a well-known graph embedding framework. We first construct an item graph from users' behavior history, and learn the embeddings of all items in the graph. The item embeddings are employed to compute pairwise similarities between all items, which are then used in the recommendation process. To alleviate the sparsity and cold start problems, side information is incorporated into the graph embedding framework. We propose two aggregation methods to integrate the embeddings of items and the corresponding side information. Experimental results from offline experiments show that methods incorporating side information are superior to those that do not. Further, we describe the platform upon which the embedding methods are deployed and the workflow to process the billion-scale data in Taobao. Using A/B test, we show that the online Click-Through-Rates (CTRs) are improved comparing to the previous collaborative filtering based methods widely used in Taobao, further demonstrating the effectiveness and feasibility of our proposed methods in Taobao's live production environment.

\end{abstract}



\maketitle

\section{Introduction}

\begin{figure*}[h]
      \centering
      \includegraphics[scale=0.40]{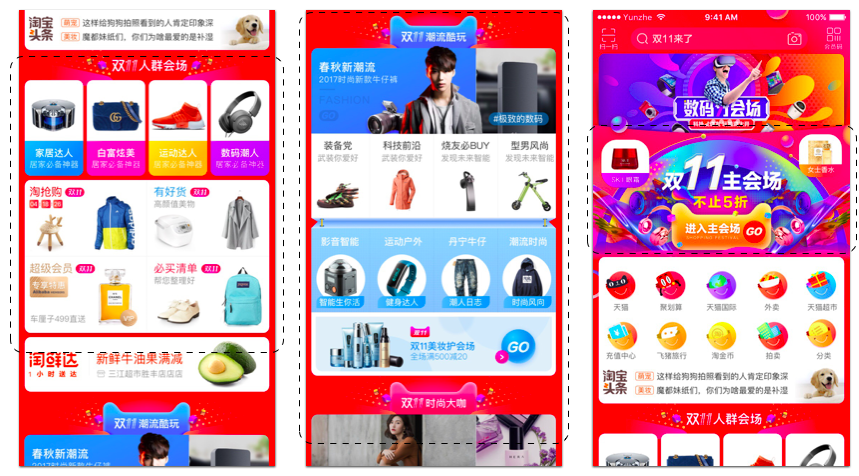}
      \caption{The areas highlighted with dashed rectangles are personalized for one billion users in Taobao. Attractive images and textual descriptions are also generated for better user experience. Note they are on Mobile Taobao App homepage, which contributes 40\% of the total recommending traffic.}
      \label{fig:intro_pic}
      \vspace{-0.2cm}
\end{figure*}

Internet technology has been continuously reshaping the business landscape, and online businesses are everywhere nowadays. Alibaba, the largest provider of online business in China, makes it possible for people or companies all over the world to do business online. With one billion users, the Gross Merchandise Volume (GMV) of Alibaba in 2017 is 3,767 billion Yuan and the revenue in 2017 is 158 billion Yuan. In the famous Double-Eleven Day, the largest online shopping festival in China, in 2017, the total amount of transactions was around 168 billion Yuan. Among all kinds of online platforms in Alibaba, Taobao\footnote{https://www.taobao.com/}, the largest online consumer-to-consumer (C2C) platform, stands out by contributing 75\% of the total traffic in Alibaba E-commerce. 

With one billion users and two billion items, i.e., commodities, in Taobao, the most critical problem is how to help users find the needed and interesting items quickly. To achieve this goal, recommendation, which aims at providing users with interesting items based on their preferences, becomes the key technology in Taobao. For example, the homepage on Mobile Taobao App (see Figure~\ref{fig:intro_pic}), which are generated based on users' past behaviors with recommendation techniques, contributes 40\% of the total recommending traffic. Furthermore, recommendation contributes the majority of both revenues and traffic in Taobao. In short, recommendation has become the vital engine of GMV and revenues of Taobao and Alibaba. Despite the success of various recommendation methods in academia and industry, e.g., collaborative filtering (CF)~\cite{herlocker1999algorithmic,sarwar2001item,linden2003amazon}, content-based methods~\cite{balabanovic1997fab}, and deep learning based methods~\cite{wang2015collaborative,cheng2016wide,covington2016deep}, the problems facing these methods become more severe in Taobao because of the billion-scale of users and items.

There are three major technical challenges facing RS in Taobao:
\begin{itemize}
	\item \textbf{Scalability}: Despite the fact that many existing recommendation approaches work well on smaller scale datasets, i.e., millions of users and items, they fail on the much larger scale dataset in Taobao, i.e., one billion users and two billion items. 
    \item \textbf{Sparsity}: Due to the fact that users tend to interact with only a small number of items, it is extremely difficult to train an accurate recommending model, especially for users or items with quite a small number of interactions. It is usually referred to as the ``sparsity'' problem.
    \item \textbf{Cold Start}: In Taobao, millions of new items are continuously uploaded each hour. There are no user behaviors for these items. It is challenging to process these items or predict the preferences of users for these items, which is the so-called ``cold start'' problem.
 \end{itemize}

To address these challenges in Taobao, we design a two-stage recommending framework in Taobao's technology platform. The first stage is \textbf{matching}, and the second is \textbf{ranking}. In the matching stage, we generate a candidate set of similar items for each item users have interacted with, and then in the ranking stage, we train a deep neural net model, which ranks the candidate items for each user according to his or her preferences. Due to the aforementioned challenges, in both stages we have to face different unique problems. Besides, the goal of each stage is different, leading to separate technical solutions. 

In this paper, we focus on how to address the challenges in the matching stage, where the core task is the computation of pairwise similarities between all items based on users' behaviors. After the pairwise similarities of items are obtained, we can generate a candidate set of items for further personalization in the ranking stage. To achieve this, we propose to construct an item graph from users' behavior history and then apply the state-of-art graph embedding methods~\cite{perozzi2014deepwalk,tang2015line,grover2016node2vec} to learn the embedding of each item, dubbed Base Graph Embedding (BGE). In this way, we can generate the candidate set of items based on the similarities computed from the dot product of the embedding vectors of items. Note that in previous works, CF based methods are used to compute these similarities. However, CF based methods only consider the co-occurrence of items in users' behavior history~\cite{herlocker1999algorithmic,sarwar2001item,linden2003amazon}. In our work, using random walk in the item graph, we can capture higher-order similarities between items. Thus, it is superior to CF based methods. However, it's still a challenge to learn accurate embeddings of items with few or even no interactions. To alleviate this problem, we propose to use side information to enhance the embedding procedure, dubbed Graph Embedding with Side information (GES). For example, items belong to the same category or brand should be closer in the embedding space. In this way, we can obtain accurate embeddings of items with few or even no interactions. However, in Taobao, there are hundreds of types of side information, like category, brand, or price, etc., and it is intuitive that different side information should contribute differently to learning the embeddings of items. Thus, we further propose a weighting mechanism when learning the embedding with side information, dubbed Enhanced Graph Embedding with Side information (EGES). 

In summary, there are three important parts in the matching stage:
\begin{enumerate}
\item Based on years of practical experience in Taobao, we design an effective heuristic method to construct the item graph from the behavior history of one billion users in Taobao.

\item We propose three embedding methods, BGE, GES, and EGES, to learn embeddings of two billion items in Taobao. We conduct offline experiments to demonstrate the effectiveness of GES and EGES comparing to BGE and other embedding methods.

\item To deploy the proposed methods for billion-scale users and items in Taobao, we build the graph embedding systems on the XTensorflow (XTF) platform constructed by our team. We show that the proposed framework significantly improves recommending performance on the Mobile Taobao App, while satisfying the demand of training efficiency and instant response of service even on the Double-Eleven Day.

\end{enumerate}

The rest of the paper is organized as follows. In Section~\ref{sec:framework}, we elaborate on the three proposed embedding methods. Offline and online experimental results are presented in Section~\ref{sec:exp}. We introduce the deployment of the system in Taobao in Section~\ref{sec:dep}, and review the related work in Section~\ref{sec:rel}. We conclude our work in Section~\ref{sec:conclusion}.

\section{Framework}
\label{sec:framework}
In this section, we first introduce the basics of graph embedding, and then elaborate on how we construct the item graph from users' behavior history. Finally, we study the proposed methods to learn the embeddings of items in Taobao.

\subsection{Preliminaries}
In this section, we give an overview of graph embedding and one of the most popular methods, DeepWalk~\cite{perozzi2014deepwalk}, based on which we propose our graph embedding methods in the matching stage. Given a graph $\cG = (\cV, \cE)$, where $\cV$ and $\cE$ represent the node set and the edge set, respectively. Graph embedding is to learn a low-dimensional representation for each node $v \in \cV$ in the space $\mathbb{R}^d$, where $d \ll |\cV|$. In other words, our goal is to learn a mapping function $\Phi: \cV \rightarrow \mathbb{R}^d$, i.e., representing each node in $\cV$ as a $d$-dimensional vector.

In~\cite{mikolov2013distributed,mikolov2013efficient}, \textit{word2vec} was proposed to learn the embedding of each word in a corpus. Inspired by word2vec, Perozzi\etal proposed DeepWalk to learn the embedding of each node in a graph~\cite{perozzi2014deepwalk}. They first generate sequences of nodes by running random walk in the graph, and then apply the Skip-Gram algorithm to learn the representation of each node in the graph. To preserve the topological structure of the graph, they need to solve the following optimization problem:
\begin{equation}
\label{eq:word2vec}
\minimize_{\Phi} \sum_{v \in \cV}\sum_{c \in N(v)} -\log Pr(c|\Phi(v))\,,
\end{equation}
where $N(v)$ is the neighborhood of node $v$, which can be defined as nodes within one or two hops from $v$. $Pr(c|\Phi(v))$ defines the conditional probability of having a context node $c$ given a node $v$.

In the rest of this section, we first present how we construct the item graph from users' behaviors, and then propose the graph embedding methods based on DeepWalk for generating low-dimensional representation for two billion items in Taobao.

\subsection{Construction of Item Graph from Users' Behaviors}
\label{sec:graph-con}

\begin{figure*}
	\centering
	\subfloat[Users' behavior sequences.\label{fig:ubs}]{\includegraphics[scale=0.28]{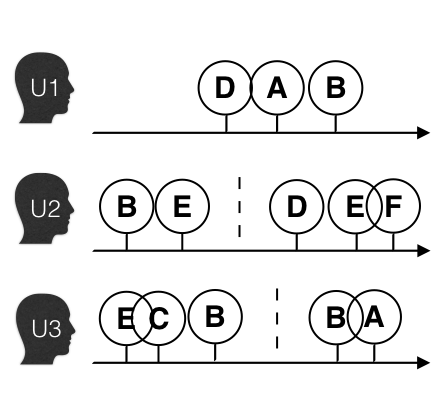}}
	\subfloat[Item graph construction.]{\includegraphics[scale=0.28]{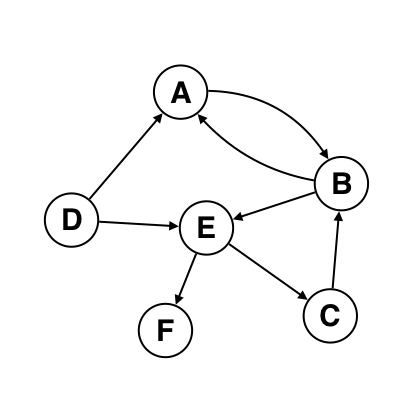}}
	\subfloat[Random walk generation.]{\includegraphics[scale=0.28]{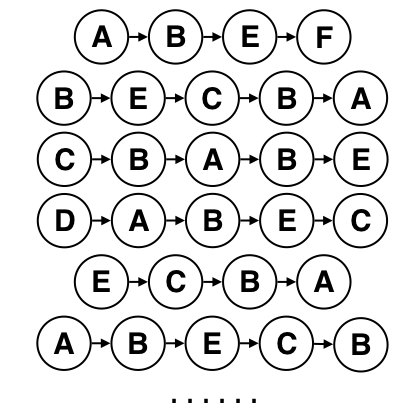}}
	\subfloat[Embedding with Skip-Gram.]{\includegraphics[scale=0.28]{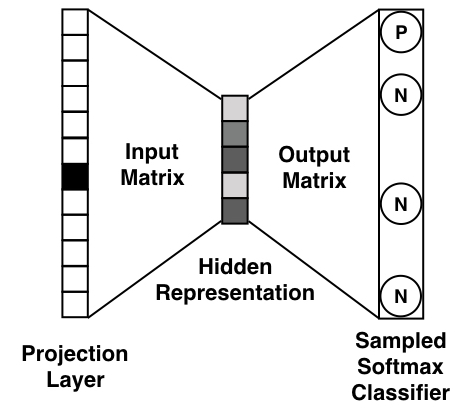}}
	\caption{Overview of graph embedding in Taobao: (a) Users' behavior sequences: One session for user u1, two sessions for user u2 and u3; these sequences are used to construct the item graph; (b) The weighted directed item graph $\cG = (\cV, \cE)$; (c) The sequences generated by random walk in the item graph; (d) Embedding with Skip-Gram.}
	\label{fig:s1}
\end{figure*}

In this section, we elaborate on the construction of the item graph from users' behaviors. In reality, a user's behaviors in Taobao tend to be sequential as shown in Figure~\ref{fig:s1} (a). Previous CF based methods only consider the co-occurrence of items, but ignore the sequential information, which can reflect users' preferences more precisely. However, it is not possible to use the whole history of a user because 1) the computational and space cost will be too expensive with so many entries; 2) a user's interests tend to drift with time. Therefore, in practice, we set a time window and only choose users' behaviors within the window. This is called session-based users' behaviors. Empirically, the duration of the time window is one hour.

After we obtain the session-based users' behaviors, two items are connected by a directed edge if they occur consecutively, e.g., in Figure~\ref{fig:s1} (b) item D and item A are connected because user $u_1$ accessed item D and A consecutively as shown in Figure~\ref{fig:s1} (a). By utilizing the collaborative behaviors of all users in Taobao, we assign a weight to each edge $e_{ij}$ based on the total number of occurrences of the two connected items in all users' behaviors. Specifically, the weight of the edge is equal to the frequency of item $i$ transiting to item $j$ in the whole users' behavior history. In this way, the constructed item graph can represent the similarity between different items based on all of the users' behaviors in Taobao.

In practice, before we extract users' behavior sequences, we need to filter out invalid data and abnormal behaviors to eliminate noise for our proposed methods.
Currently, the following behaviors are regarded as noise in our system:
\begin{itemize}
	\item If the duration of the stay after a click is less than one second, the click may be unintentional and needs to be removed.
	\item There are some ``over-active'' users in Taobao, who are actually spam users. According to our long-term observations in Taobao, if a single user bought 1,000 items or his/her total number of clicks is larger than 3,500 in less than three months, it is very likely that the user is a spam user. We need to filter out the behaviors of these users.
	\item Retailers in Taobao keep updating the details of a commodity. In the extreme case, a commodity can become a totally different item for the same identifier in Taobao after a long sequence of updates. Thus, we remove the item related to the identifier.
\end{itemize}

\subsection{Base Graph Embedding}
\label{section:ge}

After we obtain the weighted directed item graph, denoted as $\cG = (\cV, \cE)$, we adopt DeepWalk to learn the embedding of each node in $\cG$. Let $\bM$ denote the adjacency matrix of $\cG$ and $\bM_{ij}$ the weight of the edge from node $i$ pointing to node $j$.
We first generate node sequences based on random walk and then run the Skip-Gram algorithm on the sequences. The transition probability of random walk is defined as
\begin{equation}
P(v_j|v_i) = \begin{cases}
\frac{\bM_{ij}}{\sum\limits_{j \in N_+(v_i)}\bM_{ij}}\,,\qquad & v_j \in N_+(v_i)\,,\\
0\,,\qquad &e_{ij} \notin \cE\,,\\
\end{cases}
\label{eq:transit-prob}
\end{equation}
where $N_+(v_i)$ represents the set of outlink neighbors, i.e. there are edges from $v_i$ pointing to all of the nodes in $N_+(v_i)$. By running random walk, we can generate a number of sequences as shown in Figure~\ref{fig:s1} (c).

Then we apply the Skip-Gram algorithm~\cite{mikolov2013distributed,mikolov2013efficient} to learn the embeddings, which maximizes the co-occurrence probability of two nodes in the obtained sequences. This leads to the following optimization problem:
\begin{equation}
\minimize_{\Phi} -\log Pr\big(\{v_{i-w},\dotsm,v_{i+w}\} \backslash v_i|\Phi(v_i)\big)\,,
\label{eq:skip-gram-obj}
\end{equation}
where $w$ is the window size of the context nodes in the sequences. Using the independence assumption, we have
\begin{equation}
Pr\big(\{v_{i-w},\dotsm,v_{i+w}\} \backslash v_i|\Phi(v_i)\big) = \prod_{j=i-w,j \neq i}^{i+w} Pr\big(v_j | \Phi(v_i)\big)\,.
\end{equation}

Applying negative sampling~\cite{mikolov2013distributed,mikolov2013efficient}, Eq.~\eqref{eq:skip-gram-obj} can be transformed into
\begin{equation}
\minimize_{\Phi} \log \sigma\big(\Phi(v_j)^T\Phi(v_i)\big) + \sum_{t \in N(v_i)'} \log\sigma\big(-\Phi(v_t)^T\Phi(v_i)\big)\,.
\end{equation}
where $N(v_i)'$ is the negative samples for $v_i$, and $\sigma()$ is the sigmoid function $\sigma(x) = \frac{1}{1+e^{-x}}$. Empirically, the larger $|N(v_i)'|$ is, the better the obtained results.

\subsection{Graph Embedding with Side Information}
By applying the embedding method in Section~\ref{section:ge}, we can learn embeddings of all items in Taobao to capture higher-order similarities in users' behavior sequences, which are ignored by previous CF based methods. However, it is still challenging to learn accurate embeddings for ``cold-start'' items, i.e., those with no interactions of users.

\begin{figure}[h]
	\centering
	\includegraphics[width=0.95\columnwidth]{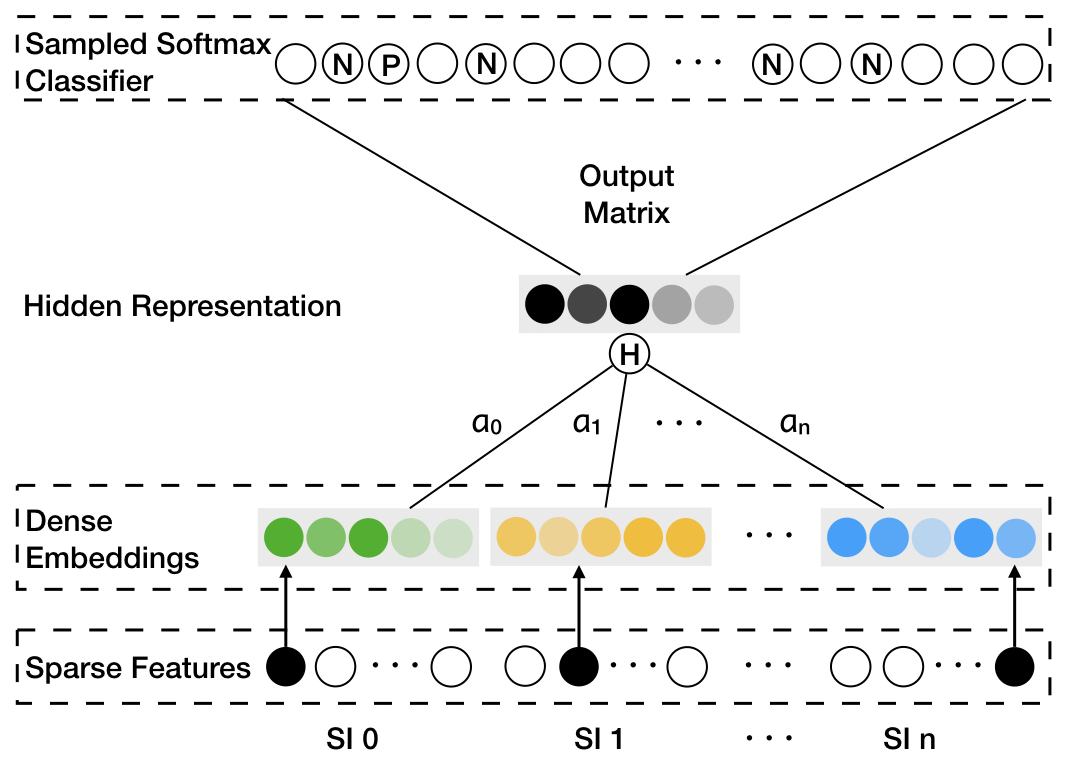}
	\caption{The general framework of GES and EGES. SI denotes the side information, and ``SI 0'' represents the item itself. In practice, 1) Sparse features tend to be one-hot-encoder vectors for items and different SIs. 2) Dense embeddings are the representation of items and the corresponding SI. 3) The hidden representation is the aggregation embedding of an item and its corresponding SI.}
	\label{fig:ges}
\end{figure}

To address the cold-start problem, we propose to enhance BGE using side information attached to cold-start items. In the context of RSs in e-commerce, side information refers to the category, shop, price, etc., of an item, which are widely used as key features in the ranking stage but rarely applied in the matching stage. We can alleviate the cold-start problem by incorporating side information in graph embedding. For example, two hoodies (same category) from UNIQLO (same shop) may look alike, and a person who likes Nikon lens may also has an interest in Canon Camera (similar category and similar brand). It means that items with similar side information should be closer in the embedding space. Based on this assumption, we propose the GES method as illustrated in Figure~\ref{fig:ges}.

For the sake of clarity, we modify slightly the notations. We use $\bW$ to denote the embedding matrix of items or side information. Specifically, $\bW^0_v$ denotes the embedding of item $v$, and $\bW^s_v$ denotes the embedding of the $s$-th type of side information attached to item $v$. Then, for item $v$ with $n$ types of side information, we have $n+1$ vectors $\bW^0_v,\dotsm,...\bW^n_v \in \mathbb{R}^d$, where $d$ is the embedding dimension. Note that the dimensions of the embeddings of items and side information are empirically set to the same value.

As shown in Figure~\ref{fig:ges}, to incorporate side information, we concatenate the $n+1$ embedding vectors for item $v$ and add a layer with average-pooling operation to aggregate all of the embeddings related to item $v$, which is
\begin{equation}
\bH_v=\frac{1}{n+1} \sum_{s=0}^n{\bW_v^s}\,,
\label{eq:avg-agg}
\end{equation}
where $\bH_v$ is the aggregated embeddings of item $v$. In this way, we incorporate side information in such a way that items with similar side information will be closer in the embedding space. This results in more accurate embeddings of cold-start items and improves the offline and online performance (see Section~\ref{sec:exp}).

\subsection{Enhanced Graph Embedding with Side Information}
\label{sec:ages}

Despite the performance gain of GES, a problem remains when integrating different kinds of side information in the embedding procedure. In Eq.~\eqref{eq:avg-agg}, the assumption is that different kinds of side information contribute equally to the final embedding, which does not reflect the reality. For example, a user who has bought an iPhone tends to view 
Macbook or iPad because of the brand ``Apple'', while a user may buy clothes of different brands in the same shop in Taobao for convenience and lower price. Therefore, different kinds of side information contribute differently to the co-occurrence of items in users' behaviors.

To address this problem, we propose the EGES method to aggregate different types of side information. The framework is the same to GES (see Figure~\ref{fig:ges}). The idea is that different types of side information have different contributions when their embeddings are aggregated. Hence, we propose a weighted average layer to aggregate the embeddings of the side information related to the items. Given an item $v$, let $\bA \in \mathbb{R}^{|V| \times (n+1)}$ be the weight matrix and the entry $\bA_{ij}$ the weight of the $j$-th type of side information of the $i$-th item. Note that $\bA_{*0}$, i.e., the first column of $\bA$, denotes the weight of item $v$ itself. For simplicity, we use $a_v^s$ to denote the weight of the $s$-th type of side information of item $v$ with $a_v^0$ denoting the weight of item $v$ itself. The weighted average layer combining different side information is defined in the following:
\begin{equation}
\label{eq:attention}
\bH_v=\frac{\sum_{j=0}^n e^{a_v^j}\bW_v^j}{\sum_{j=0}^n e^{a_v^j}}\,,
\end{equation}
where we use $e^{a_v^j}$ instead of $a_v^j$ to ensure that the contribution of each side information is greater than $0$, and $\sum_{j=0}^n e^{a_v^j}$ is used to normalize the weights related to the embeddings of different side information.

For node $v$ and its context node $u$ in the training data, we use $\bZ_u \in \mathbb{R}^d$ to represent its embedding and $y$ to denote the label. Then, the objective function of EGES becomes
\begin{align}
\label{eq:ages-loss}
\mathcal{L}(v,u,y)= -[ylog(\sigma(\bH_v^T\bZ_u)) + (1-y)log(1-\sigma(\bH_v^T\bZ_u))]\,.
\end{align}

To solve it, the gradients are derived in the following:

\begin{equation}
\label{eq:partial_z}
\frac{\partial \mathcal{L}}{\partial{\bZ_u}} 
=
(\sigma(\bH_v^T\bZ_u)-y)\bH_v\,.
\end{equation}

For $s$-th side information
\begin{align}
\frac{\partial \mathcal{L}}{\partial a_v^s}
&=
\label{eq:partial_a}
\frac{\partial \mathcal{L}}{\partial \bH_v}  
\frac{\partial \bH_v}{\partial a_v^s} \notag\\
&=
(\sigma(\bH_v^T\bZ_u)-y)\bZ_u
\frac{(\sum_{j=0}^n e^{a_v^j})e^{a_v^s}\bW_v^s - e^{a_v^s}\sum_{j=0}^n e^{a_v^j}\bW_v^j}{(\sum_{j=0}^n e^{a_v^j})^2}\,,\\
\frac{\partial \mathcal{L}}{\partial \bW_v^s} 
&=
\frac{\partial \mathcal{L}}{\partial \bH_v}  
\frac{\partial \bH_v}{\partial \bW_v^s} \nonumber\\
&= 
\frac{e^{a_v^s}}{\sum_{j=0}^n e^{a_v^j}}
(\sigma(\bH_v^T\bZ_u)-y)\bZ_u\,.
\label{eq:partial_w}
\end{align}

The pseudo code of EGES is listed in Algorithm \ref{alg:Framwork}, and the pseudo code of the weighted Skip-Gram updater is shown in Algorithm~\ref{alg:wsg}. The final hidden representation of each item is computed by Eq.~\eqref{eq:attention}.

\begin{algorithm}[htb]
	\caption{Framework of \textbf{EGES}.}
	\label{alg:Framwork}
	\begin{algorithmic}[1]
		\INPUT
		\Statex The item graph $\cG = (\cV,\cE)$, side information $S$, number of walks per node $w$, walk length $l$, Skip-Gram window size $k$, number of negatives samples $\#ns$, embedding dimension $d$;
		\OUTPUT
		\Statex The item \& side-information embeddings $\bW^0,\ldots,\bW^n$
		\Statex Weight matrix $\textbf{A}$;
		\State Initialize $\bW^0,\ldots,\bW^n,\textbf{A}$;
		\For{$i=1 \rightarrow w$}
			\For{$v \in \cV$}
			\State $SEQ$ = RandomWalk($\cG$,$v$,$l$); (Eq.~\eqref{eq:transit-prob})
			\State \textbf{WeightedSkipGram}($\bW^0,\ldots,\bW^n,\textbf{A},k,\#ns ,l,SEQ$); 
			\EndFor
		\EndFor
		\State \Return $\bW^0,\ldots,\bW^n,\textbf{A}$;
	\end{algorithmic}
\end{algorithm}

\begin{algorithm}[htb]
	\caption{Weighted Skip-Gram.}
	\label{alg:wsg}
	\begin{algorithmic}[1]
		\Function{WeightedSkipGram}{$\textbf{W}^0,\dotsm,\textbf{W}^n,\textbf{A},k,\#ns ,l,SEQ$}
		\For{$i=1 \rightarrow l$}
		\State $v=SEQ[i]$;
		\For{$j = max(0,i-k) \rightarrow min(i+k,l)$ \& $j \neq i$ }
		\State $u=SEQ[j]$
		\State Update($v,u,1$)
		\For{$t = 0 \rightarrow \#ns$}
		\State $u=NegativeSampling(V)$
		\State Update($v,u,0$)
		\EndFor
		\EndFor
		\EndFor
		\EndFunction
		\Statex
		\Function{Update}{$v,u,y$}
		\State $ \bZ_u^{new} = \bZ_u^{old} - \eta \cdot \frac{\partial}{\partial \bZ_u}\mathcal{L}$; (Eq.~\eqref{eq:partial_z})
		\For{ $s=0 \rightarrow n $}
		\State $ a_v^{s^{new}} = a_v^s{^{old}} - \eta \cdot \frac{\partial\mathcal{L}}{\partial a_v^s}$; (Eq.~\eqref{eq:partial_a})
		\State $ \bW_v^{s^{new}} = \bW_v^{s^{old}} - \eta \cdot \frac{\partial\mathcal{L}}{\partial \bW_v^s}$;  (Eq.~\eqref{eq:partial_w})
		\EndFor
		\EndFunction
	\end{algorithmic}
\end{algorithm}

\section{Experiments}
\label{sec:exp}
In this section, we conduct extensive experiments to demonstrate the effectiveness of our proposed methods. First, we evaluate the methods by the link prediction task, and then report the online experimental results on Mobile Taobao App. Finally, we present some real-world cases to give insight into the proposed methods in Taobao.

\subsection{Offline Evaluation}
\label{sec:exp-offline}
\textbf{Link Prediction.}
The link prediction task is used in the offline experiments because it is a fundamental problem in networks. Given a network with some edges removed, the link prediction task is to predict the occurrence of links.
Following similar experimental settings in~\cite{zhou2017scalable}, 1/3 of the edges are randomly chosen and removed as ground truth in the test set, and the remaining graph is taken as the training set. The same number of node pairs in the test data with no edges connecting them are randomly chosen as negative samples in the test set. To evaluate the performance of link prediction, the Area Under Curve (AUC) score is adopted as the performance metric.

\textbf{Dataset.} We use two datasets for the link prediction task. The first is Amazon Electronics\footnote{http://jmcauley.ucsd.edu/data/amazon/} provided by~\cite{mcauley2015image}, denoted as Amazon. The second is extracted from Mobile Taobao App, denoted as Taobao. Both of these two datasets include different types of side information. For the Amazon dataset, the item graph is constructed from ``co-purchasing'' relations (denoted as \textit{also\_bought} in the provided data), and three types of side information are used, i.e., category, sub-category and brand. For the Taobao dataset, the item graph is constructed according to Section~\ref{sec:graph-con}. Note that, for the sake of efficiency and effectiveness, twelve types of side information are used in Taobao's live production, including retailer, brand, purchase level, age, gender, style, etc. These types of side information have been demonstrated to be useful according to years of practical experience in Taobao. The statistics of the two datasets are shown in Table~\ref{tb:statistics}. We can see that the sparsity of the two datasets are greater than 99\%.

\begin{table}[]
	\centering
	\caption{Statistics of the two datasets. \#SI denotes the number of types of side information. Sparsity is computed according to $1 - \frac{\#Edges}{\#Nodes \times (\#Nodes - 1)}$.}
	\label{tb:statistics}
	\begin{tabular}{c|cccc}
		\hline
		Dataset      & \#Nodes & \#Edges  & \#SI  & Sparsity(\%)     \\ \hline
		Amazon  & 300,150  & 3,740,196  & 3       &  0.9958\\ \hline
		Taobao       & 2,632,379 & 44,997,887 & 12   &  0.9994   \\ \hline
	\end{tabular}
\end{table}

\textbf{Comparing Methods.} Experiments are conducted to compare four methods: BGE, LINE, GES, and EGES. LINE was proposed in~\cite{tang2015line}, which captures the first-order and second-order proximity in graph embedding. We use the implementation provided by the authors\footnote{https://github.com/tangjianpku/LINE}, and run it using first-order and second-order proximity, which are denoted, respectively, as LINE(1st) and LINE(2nd). We implement the other three methods. The emdedding dimension of all the methods is set to 160. For our BGE, GES and EGES, the length of random walk is $10$, the number of walks per node is 20, and the context window is $5$.

\begin{table}[]
	\centering
	\caption{AUCs of different methods on the two datasets. Percentages in the brackets are the improvements of AUC comparing to BGE.}
	\label{tb:auc}
	\begin{tabular}{c|cc}
		\hline
		Dataset            & Amazon        & Taobao  \\ \hline
		BGE           & 0.9327            & 0.8797 \\ \hline
		LINE(1st)         & 0.9554(+2.43\%)           & 0.9100(+3.44\%) \\
		\hline
		LINE(2nd)        & 0.8664(-7.65\%)           & 0.9411(+6.98\%) \\
		\hline
		GES                & 0.9575(+2.66\%)            & 0.9704(+10.1\%) \\ \hline
		EGES               & 0.9700(+4.00\%)            & 0.9746(+10.8\%) \\ \hline
	\end{tabular}
\end{table}

\textbf{Results Analysis.} The results are shown in Table~\ref{tb:auc}. We can see that GES and EGES outperform BGE, LINE(1st) and LINE(2st) in terms of AUC on both datasets. This demonstrates the effectiveness of the proposed methods. In other words, the sparsity problem is alleviated by incorporating side information. When comparing the improvements on Amazon and Taobao, we can see that the performance gain is more significant on Taobao dataset. We attribute this to the larger number of types of effective and informative side information used on Taobao dataset. When comparing GES and EGES, we can see that the performance gain on Amazon is lager than that on Taobao. It may be due to the fact that the performance on Taobao is already very good, i.e., 0.97. Thus, the improvement of EGES is not prominent. On Amazon dataset, EGES outperforms GES significantly in terms of AUC. Based on these results, we can observe that incorporating side information can be very useful for graph embedding, and the accuracy can be further improved by weighted aggregation of the embeddings of various side information.

\begin{figure}[h]
	\centering
	\includegraphics[width=0.95\columnwidth]{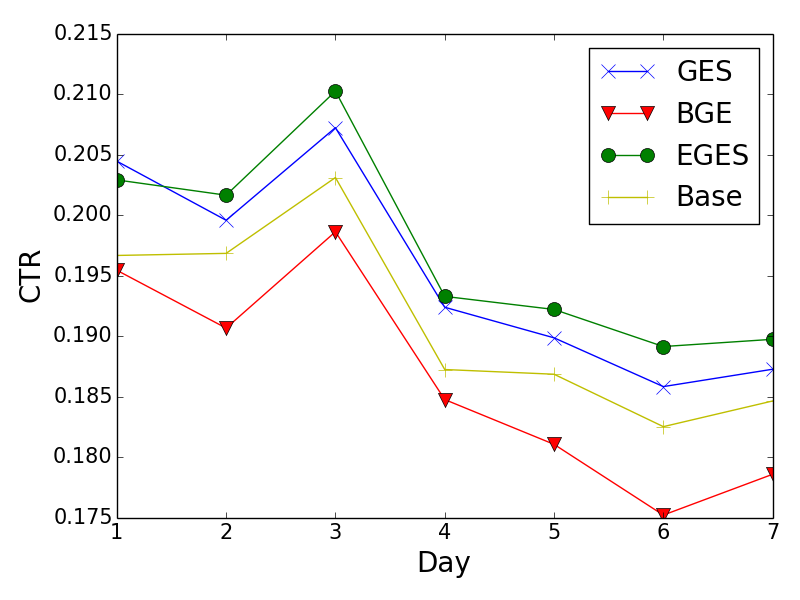}
	\caption{Online CTRs of different methods in seven days in November 2017.}
	\label{fig:ctr}
\end{figure}

\subsection{Online A/B Test}
\label{sec:exp-online}

We conduct online experiments in an A/B testing framework. The experimental goal is Click-Through-Rate (CTR) on the homepage of Mobile Taobao App. We implement the above graph embedding methods and then generate a number of similar items for each item as recommendation candidates. The final recommending results on the homepage in Taobao (see Figure~\ref{fig:intro_pic}) is generated by the ranking engine, which is implemented based on a deep neural network model. We use the same method to rank the candidate items in the experiment. As mentioned above, the quality of the similar items directly affects the recommending results. Therefore, the recommending performance, i.e., CTR, can represent the effectiveness of different methods in the matching stage. We deploy the four methods in an A/B test framework and the results of seven days in November 2017 are shown in Figure~\ref{fig:ctr}. Note that ``Base'' represents an item-based CF method which has been widely used in Taobao before graph embedding methods was deployed. It calculates the similarity between two items according to item co-occurrence and user voting weight. The similarity measurement is well-tuned and suitable for Taobao's business. 

From Figure~\ref{fig:ctr}, we can see that EGES and GES outperform BGE and Base consistently in terms of CTR, which demonstrates the effectiveness of the incorporation of side information in graph embedding. Further, the CTR of Base is larger than that of BGE. It means that well-tuned CF based methods can beat simple embedding method because a large number of hand-crafted heuristic strategies have been exploited in practice. On the other hand, EGES outperforms GES consistently, which aligns with the results in the offline experimental results in Section~\ref{sec:exp-offline}. It further demonstrates that weighted aggregation of side information is better than average aggregation.

\subsection{Case Study}
In this section, we present some real-world cases in Taobao to illustrate the effectiveness of the proposed methods. The cases are examined in three aspects: 1) visualization of the embeddings by EGES, 2) cold start items, and 3) weights in EGES.

\subsubsection{Visualization}
\begin{figure*}
	\centering
	\subfloat[Visualization of sports shoes of all categories.]{\includegraphics[width=0.45\textwidth]{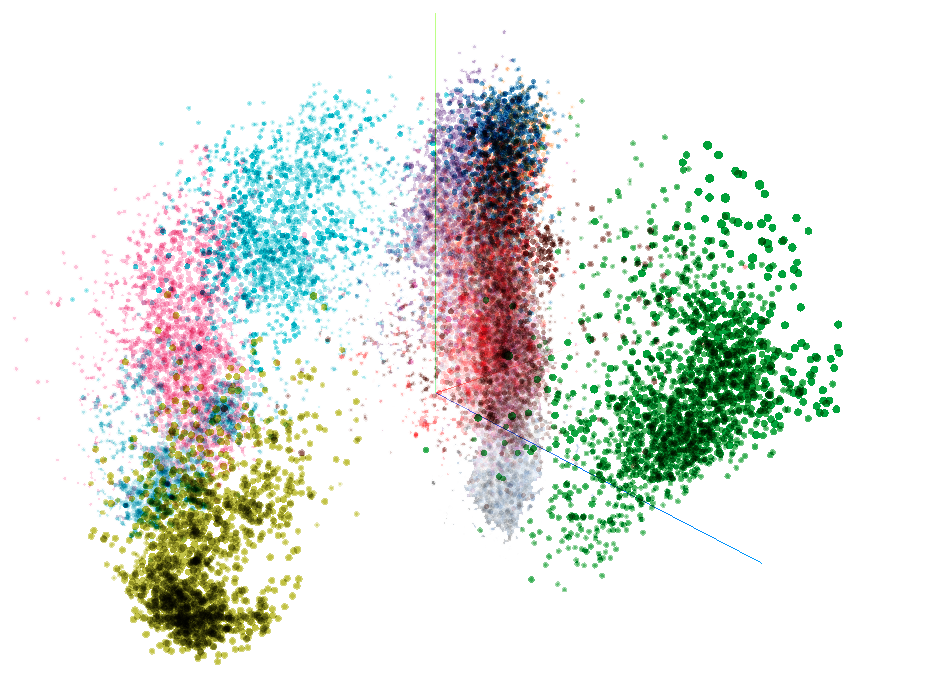}}
	\subfloat[Visualization of badminton, table tennis and football shoes. Items in gray do not belong to any of the three categories.]{\includegraphics[width=0.45\textwidth]{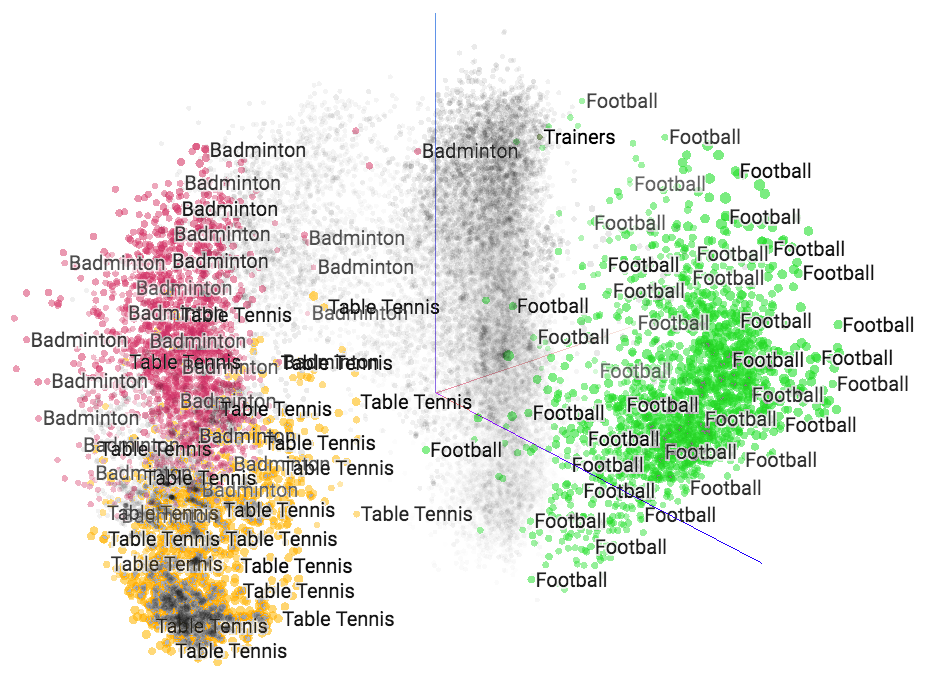}}
	\caption{Visualization of the learned embeddings of a set of randomly chosen shoes. Item embeddings are projected into a 2-D plane via principal component analysis (PCA). Different colors represent different categories. Items in the same category are grouped together.}
	\label{fig:visual}
\end{figure*}

In this part, we visualize the embeddings of items learned by EGES. We use the visualization tool provided by tensorflow\footnote{http://projector.tensorflow.org/}. The results are shown in Figure~\ref{fig:visual}. From Figure~\ref{fig:visual} (a), we can see that shoes of different categories are in separate clusters. Here one color represents one category of shoes, like badminton, table tennis, or football shoes. It demonstrates the effectiveness of the learned embeddings with incorporation of side information, i.e., items with similar side information should be closer in the embedding space. From Figure~\ref{fig:visual} (b), we further analyze the embeddings of three kinds of shoes: badminton, table tennis, and football. It is very interesting to observe that badminton and table tennis shoes are closer to each other while football shoes are farther in the embedding space. This can be explained by a phenomenon that people in China who like table tennis have much overlapping with those who like badminton. However, those who like football are quite different from those who like indoor sports, i.e., table tennis and badminton. In this sense, recommending badminton shoes to those who have viewed table tennis shoes is much better than recommending football shoes.

\begin{figure}[htb]
	\centering
	\includegraphics[width=0.95\columnwidth]{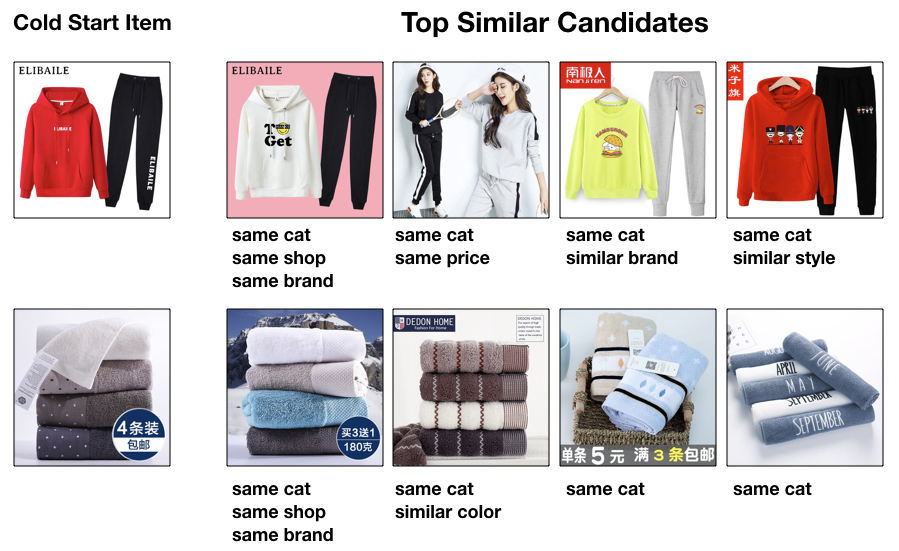}
	\caption{Similar items for cold start items. Top 4 similar items are shown. Note that ``cat'' means category.}
	\label{fig:cold_start}
\end{figure}

\subsubsection{Cold Start Items}
In this part, we show the quality of the embeddings of cold start items. For a newly updated item in Taobao, no embedding can be learned from the item graph, and previous CF based methods also fail in handling cold start items. Thus, we represent a cold start item with the average embeddings of its side information. Then, we retrieve the most similar items from the existing items based on the dot product of the embeddings of two items. The results are shown in Figure~\ref{fig:cold_start}. We can see that despite the missing of users' behaviors for the two cold start items, different side information can be utilized to learn their embeddings effectively in terms of the quality of the top similar items. In the figure, we annotate for each similar item the types of side information connected to the cold start item. We can see that the shops of the items are very informative for measuring the similarity of two items, which also aligns with the weight of each side information in the following part.

\begin{figure}[h]
	\centering
	\includegraphics[width=0.95\columnwidth]{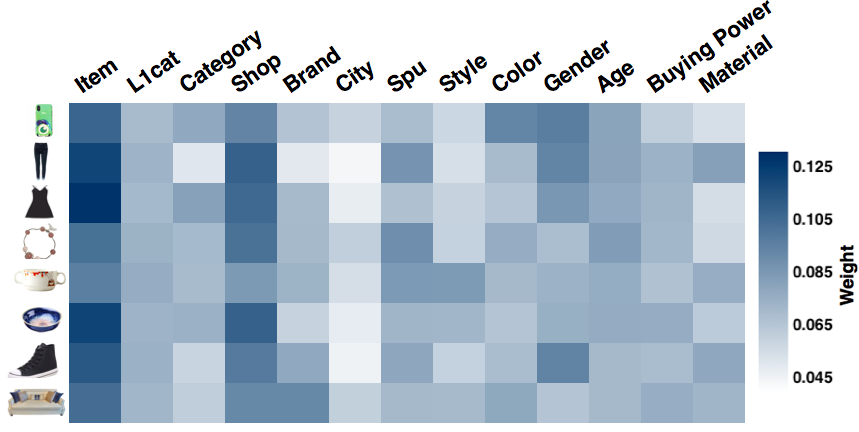}
	\caption{Weights for different side information of various items. Here ``Item'' means the embedding of an item itself.}
	\label{fig:weights}
\end{figure}

\subsubsection{Weights in EGES}
In this part, we visualize the weights of different types of side information for various items. Eight items in different categories are selected and the weights of all side information related to these items are extracted from the learned weight matrix $\bA$. The results are shown in Figure~\ref{fig:weights}, where each row records the results of one item. Several observations are worth noting: 1) The weight distributions of different items are different, which aligns with our assumption that different side information contribute differently to the final representation. 2) Among all the items, the weights of ``Item'', representing the embeddings of the item itself, are consistently larger than those of all the other side information. It confirms the intuition that the embedding of an item itself remains to be the primary source of users' behaviors whereas side information provides additional hints for inferring users' behaviors. 3) Besides ``Item'', the weights of ``Shop'' are consistently larger than those of the other side information. It aligns with users' behaviors in Taobao, that is, users tend to purchase items in the same shop for convenience and lower price.

\section{System Deployment and Operation}
\label{sec:dep}
In this section, we introduce the implementation and deployment of the proposed graph embedding methods in Taobao. We first give a high-level introduction of the whole recommending platform powering Taobao and then elaborate on the modules relevant to our embedding methods.

\begin{figure}[h]
	\centering
	\includegraphics[width=0.95\columnwidth]{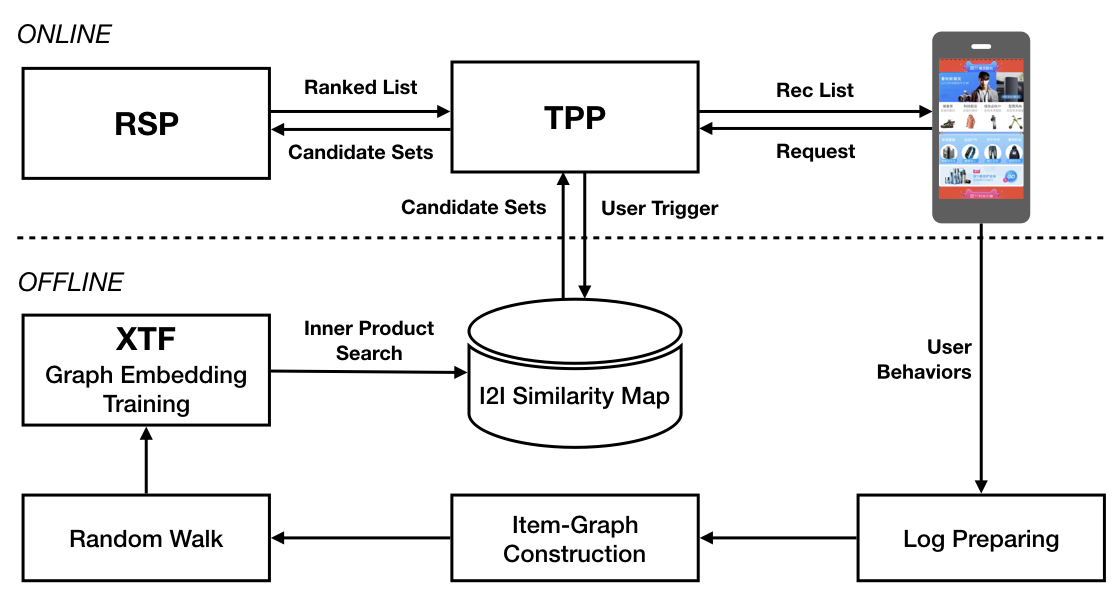}
	\caption{Architecture of the recommending platform in Taobao.}
	\label{fig:platforms}
\end{figure}

In Figure~\ref{fig:platforms}, we show the architecture of the recommending platform in Taobao. The platform consists of two subsystems: online and offline. For the online subsystem, the main components are Taobao Personality Platform (TPP) and Ranking Service Platform (RSP). A typical workflow is illustrated in the following:
\begin{itemize}
	\item When a user launches Mobile Taobao App, TPP extracts the user's latest information and retrieves a candidate set of items from the offline subsystem, which is then fed to RSP. RSP ranks the candidate set of items with a fine-tuned deep neural net model and returns the ranked results to TPP.
	\item Users' behaviors during their visits in Taobao are collected and saved as log data for the offline subsystem.
\end{itemize}

The workflow of the offline subsystem, where graph embedding methods are implemented and deployed, is described in the following:
\begin{itemize}
	\item The logs including users' behaviors are retrieved. The item graph is constructed based on the users' behaviors. In practice, we choose the logs in the recent three months. Before generating session-based users' behavior sequences, anti-spam processing is applied to the data. The remaining logs contains about 600 billion entries. Then, the item graph is constructed according to the method described in Section~\ref{sec:graph-con}.
	\item To run our graph embedding methods, two practical solutions are adopted: 1) The whole graph is split into a number of sub-graphs, which can be processed in parallel in Taobao's Open Data Processing Service (ODPS) distributed platform. There are around 50 million nodes in each subgraph. 2) To generate the random walk sequences in the graph, we use our iteration-based distributed graph framework in ODPS. The total number of generated sequences by random walk is around 150 billion.
	\item To implement the proposed embedding algorithms, 100 GPUs are used in our XTF platform. On the deployed platform, with 150 billion samples, all modules in the offline subsystem, including log retrieval, anti-spam processing, item graph construction, sequence generation by random walk, embedding, item-to-item similarity computation and map generation, can be executed in less than six hours. Thus, our recommending service can respond to users' latest behaviors in a very short time.
\end{itemize}

\section{Related Work}
\label{sec:rel}
In this section, we briefly review the related work of graph embedding, graph embedding with side information, and graph embedding for RS.

\subsection{Graph Embedding}
Graph Embedding algorithms have been proposed as a general network representation method. They have been applied to many real-world applications. In the past few years, there has been a lot of research in the field focusing on designing new embedding algorithms. These methods could be categorized into three broad categories: 1) Factorization methods such as LINE~\cite{ahmed2013distributed} try to approximately factorize the adjacency matrix and preserve both first order and second proximities; 2) Deep learning methods~\cite{wang2016structural,tu2017transnet,cao2016deep} enhance the model's ability of capturing non-linearity in graph; 3) Random walk based techniques~\cite{perozzi2014deepwalk,grover2016node2vec,dong2017metapath2vec} use random walks on graphs to obtain node representations which are extraordinary efficient and thus could be used in extremely large-scale networks. In this paper, our embedding framework is based on random walk.

\subsection{Graph Embedding with Side Information}
The above graph embedding methods only use the topological structure of the network, which suffer from the sparsity and cold start problems. In recent years, a lot of works tried to incorporate side information to enhance graph embedding methods. Most works build their tasks based on the assumption that nodes with similar side information should be closer in the embedding space. To achieve this, a joint framework was proposed to optimize the embedding objective function with a classifier function~\cite{li2016discriminative,tu2016max}. In~\cite{xie2016representation}, Xie\etal further embedded a complicated knowledge graph with the nodes in a hierarchical structure, like sub-categories, etc. Besides, textual information related to the nodes is incorporated into graph embedding~\cite{yang2015network,wang2016text,yao2017incorporating,tu2017cane}. Moreover, in~\cite{chang2015heterogeneous}, Chang\etal proposed a deep learning framework to simultaneously deal with the text and image features for heterogeneous graph embedding. In this paper, we mainly process discrete side information related to items in Taobao, such as category, brand, price, etc., and design a hidden layer to aggregate different types of side information in the embedding framework.

\subsection{Graph Embedding for RS}
RSs have been one of the most popular downstream tasks of graph embedding. With the representation in hand, various prediction models can be used to recommend. In~\cite{yu2014personalized,zhao2017meta}, embeddings of users and items are learned under the supervision of meta-path and meta-graphs, respectively, in heterogeneous information networks. Yu\etal\cite{yu2014personalized} proposed a linear model to aggregate the embeddings for recommendation while Zhao\etal\cite{zhao2017meta} proposed to apply factorization machine to the embeddings for recommendation. In~\cite{zhang2016collaborative}, Zhang\etal proposed a joint embedding framework to learn the embeddings of graph, text and images, which are used for recommendation. In~\cite{zhou2017scalable}, Zhou\etal proposed graph embedding to capture asymmetric similarities for node recommendation. In this paper, our graph embedding methods are integrated in a two-stage recommending platform. Thus, the performance of the embeddings directly affects the final recommending results.
\vspace{0.1in}

\section{Conclusion and Future Work}
\label{sec:conclusion}

Taobao's billion-scale data (one billion users and two billion items) is putting tremendous stress on its RS in terms of scalability, sparsity and cold start. In this paper, we present graph embedding based methods to address these challenges. To cope with the sparsity and cold-start problems, we propose to incorporate side information into graph embedding. Offline experiments are conducted to demonstrate the effectiveness of side information in improving recommending accuracy. Online CTRs are also reported to demonstrate the effectiveness and feasibility of our proposed methods in Taobao's live production. Real-world cases are analyzed to highlight the strength of our proposed graph embedding methods in clustering related items using users' behavior history and dealing with cold start items using side information. Finally, to address the scalability and deployment issues of our proposed solutions in Taobao, we elaborate on the the platforms for training our graph embedding methods and the overall workflow of the recommendation platform in Taobao. For future work, we will pursue two directions. The first is to utilize attention mechanism in our graph embedding methods, which can provide more flexibility to learn the weights of different side information. The second direction is to incorporate textual information into our methods to exploit the large number of reviews attached to the items in Taobao. 
\vspace{0.01in}

\section{Acknowledgments}
We would like to thank colleagues of our team - Wei Li, Qiang Liu, Yuchi Xu, Chao Li, Zhiyuan Liu, Jiaming Xu, Wen Chen and Lifeng Wang for useful discussions and supports on this work. We are grateful to our cooperative team - search engineering team. We also thank the anonymous reviewers for their valuable comments and suggestions that help improve the quality of this manuscript.

\bibliographystyle{abbrv}
\bibliography{ref} 

\begin{thebibliography}{10}

\bibitem{ahmed2013distributed}
A.~Ahmed, N.~Shervashidze, S.~Narayanamurthy, V.~Josifovski, and A.~J. Smola.
\newblock Distributed large-scale natural graph factorization.
\newblock In {\em WWW}, pages 37--48, 2013.

\bibitem{balabanovic1997fab}
M.~Balabanovi{\'c} and Y.~Shoham.
\newblock Fab: {C}ontent-based, collaborative recommendation.
\newblock {\em Communications of the ACM}, 40(3):66--72, 1997.

\bibitem{cao2016deep}
S.~Cao, W.~Lu, and Q.~Xu.
\newblock Deep neural networks for learning graph representations.
\newblock In {\em AAAI}, pages 1145--1152, 2016.

\bibitem{chang2015heterogeneous}
S.~Chang, W.~Han, J.~Tang, G.-J. Qi, C.~C. Aggarwal, and T.~S. Huang.
\newblock Heterogeneous network embedding via deep architectures.
\newblock In {\em KDD}, pages 119--128, 2015.

\bibitem{cheng2016wide}
H.-T. Cheng, L.~Koc, J.~Harmsen, T.~Shaked, T.~Chandra, H.~Aradhye,
  G.~Anderson, G.~Corrado, W.~Chai, M.~Ispir, et~al.
\newblock Wide \& deep learning for recommender systems.
\newblock Technical report, 2016.

\bibitem{covington2016deep}
P.~Covington, J.~Adams, and E.~Sargin.
\newblock Deep neural networks for youtube recommendations.
\newblock In {\em RecSys}, pages 191--198, 2016.

\bibitem{dong2017metapath2vec}
Y.~Dong, N.~V. Chawla, and A.~Swami.
\newblock metapath2vec: Scalable representation learning for heterogeneous
  networks.
\newblock In {\em KDD}, pages 135--144, 2017.

\bibitem{grover2016node2vec}
A.~Grover and J.~Leskovec.
\newblock Node2vec: {S}calable feature learning for networks.
\newblock In {\em KDD}, pages 855--864, 2016.

\bibitem{herlocker1999algorithmic}
J.~Herlocker, J.~Konstan, A.~Borchers, and J.~Riedl.
\newblock An algorithmic framework for performing collaborative filtering.
\newblock In {\em SIGIR}, pages 230--237, 1999.

\bibitem{li2016discriminative}
J.~Li, J.~Zhu, and B.~Zhang.
\newblock Discriminative deep random walk for network classification.
\newblock In {\em ACL}, volume~1, pages 1004--1013, 2016.

\bibitem{linden2003amazon}
G.~Linden, B.~Smith, and J.~York.
\newblock Amazon. com recommendations: Item-to-item collaborative filtering.
\newblock {\em IEEE Internet computing}, 7(1):76--80, 2003.

\bibitem{mcauley2015image}
J.~McAuley, C.~Targett, Q.~Shi, and A.~Van Den~Hengel.
\newblock Image-based recommendations on styles and substitutes.
\newblock In {\em SIGIR}, pages 43--52, 2015.

\bibitem{mikolov2013efficient}
T.~Mikolov, K.~Chen, G.~Corrado, and J.~Dean.
\newblock Efficient estimation of word representations in vector space.
\newblock {\em arXiv preprint arXiv:1301.3781}, 2013.

\bibitem{mikolov2013distributed}
T.~Mikolov, I.~Sutskever, K.~Chen, G.~S. Corrado, and J.~Dean.
\newblock Distributed representations of words and phrases and their
  compositionality.
\newblock In {\em NIPS}, pages 3111--3119, 2013.

\bibitem{perozzi2014deepwalk}
B.~Perozzi, R.~Al-Rfou, and S.~Skiena.
\newblock Deepwalk: Online learning of social representations.
\newblock In {\em KDD}, pages 701--710, 2014.

\bibitem{sarwar2001item}
B.~Sarwar, G.~Karypis, J.~Konstan, and J.~Riedl.
\newblock Item-based collaborative filtering recommendation algorithms.
\newblock In {\em WWW}, pages 285--295, 2001.

\bibitem{tang2015line}
J.~Tang, M.~Qu, M.~Wang, M.~Zhang, J.~Yan, and Q.~Mei.
\newblock Line: {L}arge-scale information network embedding.
\newblock In {\em WWW}, pages 1067--1077, 2015.

\bibitem{tu2017cane}
C.~Tu, H.~Liu, Z.~Liu, and M.~Sun.
\newblock Cane: Context-aware network embedding for relation modeling.
\newblock In {\em ACL}, volume~1, pages 1722--1731, 2017.

\bibitem{tu2016max}
C.~Tu, W.~Zhang, Z.~Liu, and M.~Sun.
\newblock Max-margin deepwalk: Discriminative learning of network
  representation.
\newblock In {\em IJCAI}, pages 3889--3895, 2016.

\bibitem{tu2017transnet}
C.~Tu, Z.~Zhang, Z.~Liu, and M.~Sun.
\newblock Transnet: {T}ranslation-based network representation learning for
  social relation extraction.
\newblock In {\em IJCAI}, pages 19--25, 2017.

\bibitem{wang2016structural}
D.~Wang, P.~Cui, and W.~Zhu.
\newblock Structural deep network embedding.
\newblock In {\em KDD}, pages 1225--1234, 2016.

\bibitem{wang2015collaborative}
H.~Wang, N.~Wang, and D.-Y. Yeung.
\newblock Collaborative deep learning for recommender systems.
\newblock In {\em KDD}, pages 1235--1244, 2015.

\bibitem{wang2016text}
Z.~Wang and J.-Z. Li.
\newblock Text-enhanced representation learning for knowledge graph.
\newblock In {\em IJCAI}, pages 1293--1299, 2016.

\bibitem{xie2016representation}
R.~Xie, Z.~Liu, and M.~Sun.
\newblock Representation learning of knowledge graphs with hierarchical types.
\newblock In {\em IJCAI}, pages 2965--2971, 2016.

\bibitem{yang2015network}
C.~Yang, Z.~Liu, D.~Zhao, M.~Sun, and E.~Y. Chang.
\newblock Network representation learning with rich text information.
\newblock In {\em IJCAI}, pages 2111--2117, 2015.

\bibitem{yao2017incorporating}
L.~Yao, Y.~Zhang, B.~Wei, Z.~Jin, R.~Zhang, Y.~Zhang, and Q.~Chen.
\newblock Incorporating knowledge graph embeddings into topic modeling.
\newblock In {\em AAAI}, pages 3119--3126, 2017.

\bibitem{yu2014personalized}
X.~Yu, X.~Ren, Y.~Sun, Q.~Gu, B.~Sturt, U.~Khandelwal, B.~Norick, and J.~Han.
\newblock Personalized entity recommendation: A heterogeneous information
  network approach.
\newblock In {\em WSDM}, pages 283--292, 2014.

\bibitem{zhang2016collaborative}
F.~Zhang, N.~J. Yuan, D.~Lian, X.~Xie, and W.-Y. Ma.
\newblock Collaborative knowledge base embedding for recommender systems.
\newblock In {\em KDD}, pages 353--362, 2016.

\bibitem{zhao2017meta}
H.~Zhao, Q.~Yao, J.~Li, Y.~Song, and D.~L. Lee.
\newblock Meta-graph based recommendation fusion over heterogeneous information
  networks.
\newblock In {\em KDD}, pages 635--644, 2017.

\bibitem{zhou2017scalable}
C.~Zhou, Y.~Liu, X.~Liu, Z.~Liu, and J.~Gao.
\newblock Scalable graph embedding for asymmetric proximity.
\newblock In {\em AAAI}, pages 2942--2948, 2017.

\end{thebibliography}
\end{document}